# Harmonic Generation in Metallic, GaAs-Filled Nanocavities in the Enhanced Transmission Regime at Visible and UV Wavelengths

**M.A. Vincenti[1,*], D. de Ceglia[1], V. Roppo[2] and M. Scalora[3]**

[1]*AEgis Technologies Group, 410 Jan Davis Dr., Huntsville, AL 35806, USA*

[2]*Universitat Politècnica de Catalunya, Departament de Física i Enginyeria Nuclear, Rambla Sant Nebridi, 08222 Terrassa, Spain*

[3]*Charles M. Bowden Research Center AMSRD-AMR-WS-ST, RDECOM, Redstone Arsenal, Alabama 35898,USA*

[*]*mvincenti@aegistg.com*

**Abstract:** We have conducted a theoretical study of harmonic generation from a silver grating having slits filled with GaAs. By working in the enhanced transmission regime, and by exploiting phase-locking between the pump and its harmonics, we guarantee strong field localization and enhanced harmonic generation under conditions of high absorption at visible and UV wavelengths. Silver is treated using the hydrodynamic model, which includes Coulomb and Lorentz forces, convection, electron gas pressure, plus bulk $\chi^{(3)}$ contributions. For GaAs we use nonlinear Lorentz oscillators, with characteristic $\chi^{(2)}$ and $\chi^{(3)}$ and nonlinear sources that arise from symmetry breaking and Lorentz forces. We find that: (i) electron pressure in the metal contributes to linear and nonlinear processes by shifting/reshaping the band structure; (ii) TE- and TM-polarized harmonics can be generated efficiently; (iii) the $\chi^{(2)}$ tensor of GaAs couples TE- and TM-polarized harmonics that create phase-locked pump photons having polarization orthogonal compared to incident pump photons; (iv) Fabry-Perot resonances yield more efficient harmonic generation compared to plasmonic transmission peaks, where most of the light propagates along external metal surfaces with little penetration inside its volume. We predict conversion efficiencies that range from $10^{-6}$ for second harmonic generation to $10^{-3}$ for the third harmonic signal, when pump power is $2GW/cm^2$.

## 1. Introduction

Since the first observation of enhanced optical transmission (EOT) [1] efforts have multiplied to prove that the conditions for EOT coincide with strong field localization on the metal surface and in proximity of the apertures [2-4]. Second harmonic generation (SHG) has been observed for a single aperture surrounded by grooves [5], and for arrays of sub-wavelength holes of different shapes [6] arranged in either periodic or irregular patterns [5-9]. Third harmonic generation (THG) has been demonstrated for a gold film patterned with nano-holes [10]. Metals like silver are centrosymmetric and lack a second order nonlinear term. However, they possess a relatively fast third order nonlinear response that may be among the largest of any known material ($\chi^{(3)}$~$10^{-6}$-$10^{-8}$ esu) that arises from smearing of the Fermi surface, and slower thermal contributions [11, 12], and is certain to dominate third order processes. SHG arises from a combination of symmetry breaking at the surface and from volume contributions, in part due to the magnetic Lorentz force, and calculations are performed by introducing effective surface and volume sources each having suitable weight [13-16].

Another relevant feature in harmonic generation from sub-wavelength patterned metal is the nature and topology of the apertures. Field localization and harmonic generation are significantly different whether (slits, annular structures) or not (holes) resonant modes are excited inside the aperture. The ability of slits and annular structures to support TEM-like resonant modes [17-19] indeed allows more opportunities to efficiently generate harmonic fields when the apertures are filled with nonlinear materials like $LiNbO_3$ or GaAs [20-22]. This notwithstanding, a number of important aspects are ignored in theoretical descriptions: (1) detailed dynamical contributions of the metal to SHG and THG from electron gas pressure, convection, inner core electrons and a $\chi^{(3)}$ response; (2) harmonic generation due to symmetry breaking and magnetic forces at work in GaAs; (3) phase-locking between the pump and the harmonics that allows generation in wavelength ranges below the absorption edge; (4) spectral shifts due to electron gas pressure and third order processes in metal and semiconductor sections of the grating. These elements complicate the theoretical picture, and have not been investigated in this context. At the same time, these aspects have the potential to make these structures functional in regimes where absorption is substantial and always decisive for many potential applications.

Our study of harmonic generation from metallic structures unfolds without imposing any separation between surface and volume sources by adopting the hydrodynamic model [23-27] to describe free (conduction) electrons in the metal, by making no *a priori* assumptions about charge or current distributions, and by including Coulomb (electric), Lorentz (magnetic), convective, electron gas pressure and linear and nonlinear contributions to the dielectric constant of the metal arising from inner core electrons as outlined in reference [28]. When harmonic generation is tackled in plasmonic contexts there is an understandable but critical tendency to simplify the approach by focusing on the nonlinear proprieties of the material that fills the cavity, and by ignoring any role the metal may play other than being a mere vessel [20-22]. Indeed surface sources and magnetic forces that drive bound electrons in the active material (e.g. GaAs) are routinely overlooked, along with electron gas pressure and harmonic generation that arises from the surrounding metal walls. Yet, these same elements can play a catalytic role by activating new interaction channels and should be investigated.

Another ingredient that emerges as pivotal at wavelengths below the absorption edge is a phase-locking mechanism that dominates harmonic generation in the phase-mismatched regime, and renders materials transparent at the harmonic wavelengths. In bulk materials this is exemplified by the existence of a double-peaked generated SH signal. The evidence for this phenomenon, which is produced by the homogeneous and inhomogeneous solutions of the wave equation [29-32], can be found in experimental works carried out in bulk media, where large phase and group velocity mismatches between the fundamental field (FF) and the SH waves allows the observation of two distinct SH pulses traveling at different phase and group velocities [33-35]. The homogeneous solution propagates with the phase and group velocity dictated by material dispersion, and walks off from the pump. The inhomogeneous solution is trapped by the pump and is impressed with the pump's phase and group velocity. This peculiar behavior corresponds to a phase-locking mechanism that occurs in negative index [35, 36] and absorbing materials [37, 38] as the FF co-propagates with harmonics tuned below the absorption edge [34-38]. The effect persists in a GaAs cavity, with improvements predicted and observed for SH (612nm) and TH (408nm) efficiencies [38]. There is evidence that phase-locking and transparency also occur in ranges where the dielectric constant of materials like GaP displays metallic behavior (223nm) [39]. In what follows we tune the pump at 1064nm, where GaAs is transparent, so that SH and TH fall in ranges where absorption is dominant: the components that survive experience dramatic enhancement of conversion efficiencies.

## 2. Linear response of a silver metal grating filled with GaAs

We first examined the properties of a single slit of size *a* filled with GaAs carved on an otherwise smooth silver [40] layer having thickness *w* (Fig. 1(a)). At the FF $\varepsilon_{GaAs}$(1064nm) ~12.10. In the absorbing region $\varepsilon_{GaAs}$(532nm) ~17.08+i2.86 and $\varepsilon_{GaAs}$(354nm)~8.81+ i14.36. We optimized the linear transmission using incident TM-polarized light (H-field points into the page −x-axis− in Fig. 1(a)). The slit supports TEM-like modes that exhibit field intensities more than 100 times larger than input intensities [41, 42]. We varied silver film thickness and aperture size and obtained a transmission map that reveals the resonant nature of the structure (Fig. 1(b)).

Simulations were then carried out on an array of slits 60nm wide on a 100nm-thick film. The periodicity $p$ of the array was varied from 200nm to 3200nm. In Fig. 2(a) we show the transmission for TM- (red line–square markers) and TE-polarizations (blue line–circle markers), normalized with respect to the energy that impinges on the geometrical area of the slits. In Fig. 2(b) we report the *total* transmittance as a function of wavelength for $p$=540nm, near the transmission maximum. An EOT of ~280% turns into a total transmission of 30% with a ~100nm-wide Fabry-Perot (FP) resonance centered at 1064nm. The second peak near 350nm is due mostly to the intrinsic transparency of silver. Plasmonic features (gap and resonance near 550nm) are scarcely visible and have little impact on harmonic generation compared to readily available and much more prominent cavity modes.

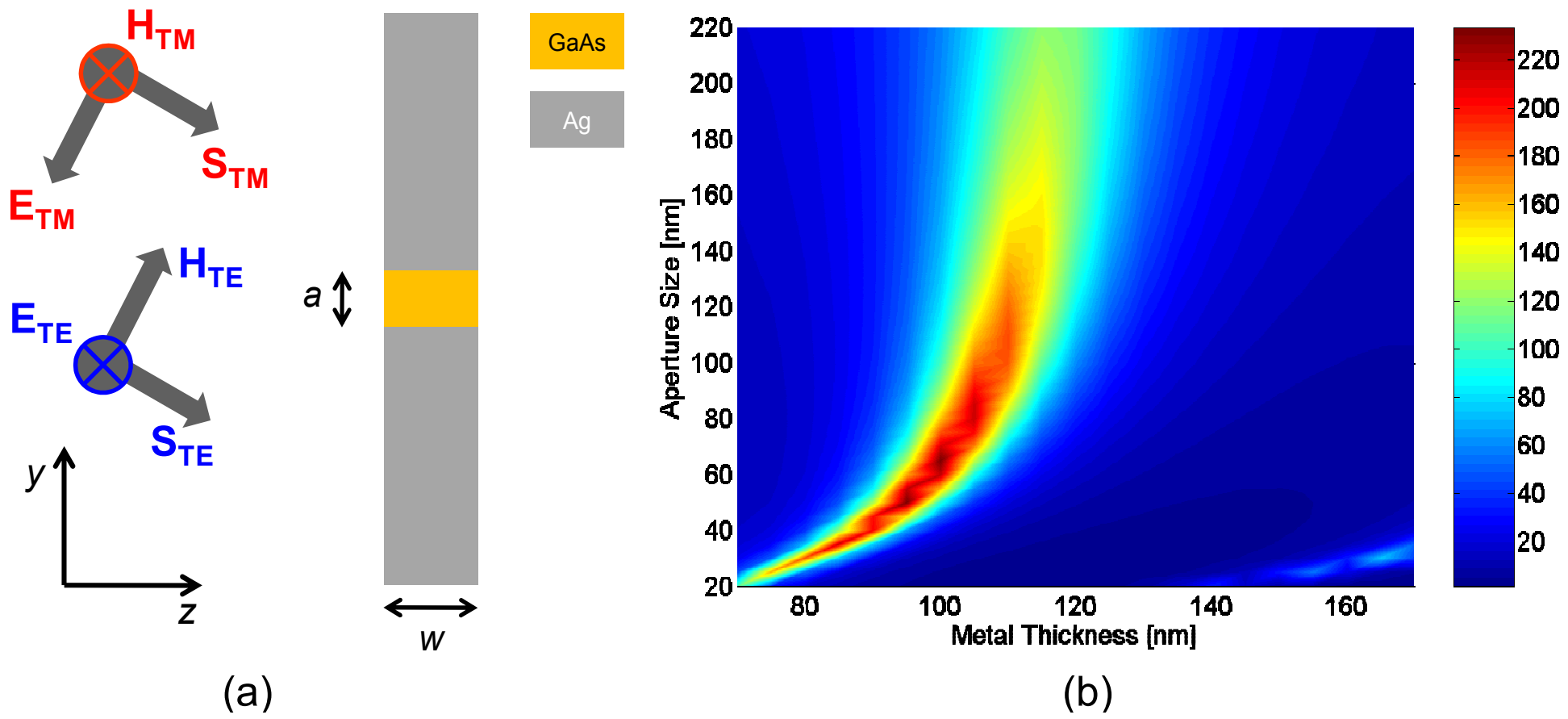

Fig.1. (a) Sketch of a single slit of size a filled with GaAs and milled in a silver film of thickness w; (b) Transmission map at λ=1064nm for a single slit on a silver substrate. We assume the fields are incident normal to the grating.

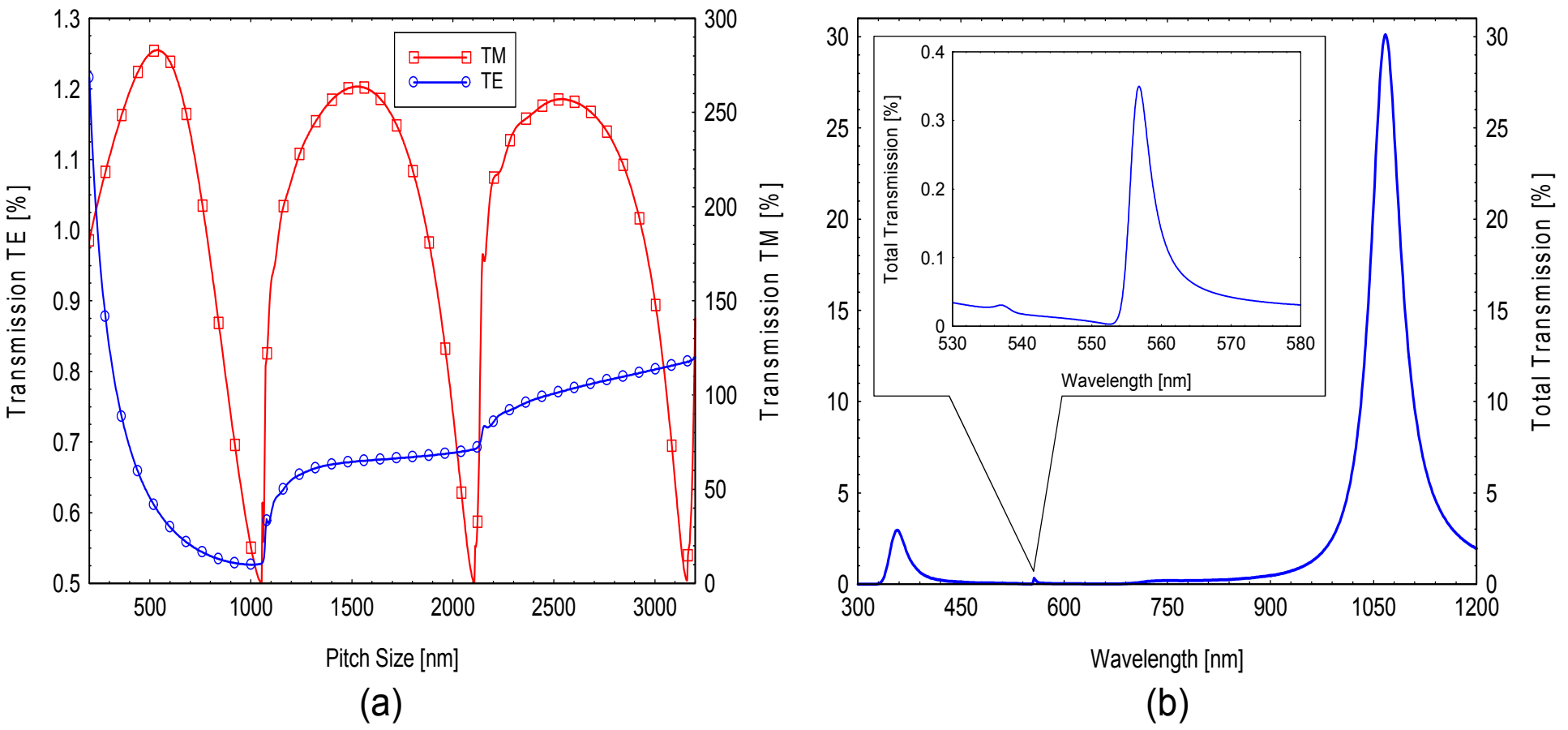

Fig.2. (a) Transmission versus pitch size $p$ at 1064nm for TM (red line–square markers, right axis) and TE (blue line–circle markers, left axis) polarizations. Transmission minima occur when $p$ matches a multiple of the plasmon wavelength; (b) Transmission vs. wavelength when $p$=540nm. Inset: plasmonic gap and resonances.

If the periodicity is chosen to be a multiple of the surface plasmon wavelength of the dielectric/metal unperturbed interface then a plasmonic band gap appears [43-46]. Slits have no cut-off for TM-polarized light, and a TEM-like resonant mode is always available, e.g. Fig. 1(b). The interference of cavity and surface modes modulates the linear transmission profile –Fig. 2(a)– and opens a gap when the bare surface plasmon wavelength matches array periodicity. Transmission values for an incident, TE-polarized pump field –Fig. 2(a)– are well below 1% for large periodicities, and approach 1% when slit-to-slit distance is small. The reason for the large difference between the two polarizations is due to the fact that resonant FP modes are not accessible to TE-polarized light, which at 1064nm are well below cut-off.

## 3. Linear and nonlinear models for Silver and GaAs

We now illustrate linear and nonlinear dynamics of a sub-wavelength patterned silver film filled with GaAs in the pulsed regime, and identify the origins of the generated signals. The model is outlined in details elsewhere [28], and here we summarize the most salient points. The hydrodynamic model describes the metal, and includes electric and magnetic forces, convection and electron gas pressure. We study a wavelength range where interband transitions are important (visible, UV) and use a Drude-Lorentz model to account for core electron contributions to the linear dielectric constant and to harmonic generation. GaAs is modeled as nonlinear Lorentz oscillators with absorption resonances at ~400nm and ~268nm [40], and all electrons are under the influence of electric and magnetic forces: we account for surface and volume contributions in all sections of the grating. Both silver and GaAs are assigned $\chi^{(2)}$ (zero for the centrosymmetric metal) and $\chi^{(3)}$ tensors typical of their crystallographic groups. Conduction electrons in the metal are described as follows [28]:

$$\begin{aligned}\ddot{\mathbf{P}}_f + \tilde{\gamma}\dot{\mathbf{P}}_f = & \frac{n_{0,f}e^2}{m_f^*}\left(\frac{\lambda_0}{c}\right)^2 \mathbf{E} - \frac{e\,\lambda_0}{m_f^* c^2}\mathbf{E}\left(\nabla\bullet\mathbf{P}_f\right) + \frac{e\,\lambda_0}{m_f^* c^2}\dot{\mathbf{P}}_f \times \mathbf{H} - \frac{1}{n_{0,f}e\lambda_0}\left[\left(\nabla\bullet\dot{\mathbf{P}}_f\right)\dot{\mathbf{P}}_f + \left(\dot{\mathbf{P}}_f\bullet\nabla\right)\dot{\mathbf{P}}_f\right] \\ & + \frac{5E_F}{3m_f^* c^2}\nabla\left(\nabla\bullet\mathbf{P}_f\right) - \frac{10}{9}\frac{E_F / m_f^* c^2}{n_{0f}e\lambda_0}\left(\nabla\bullet\mathbf{P}_f\right)\nabla\left(\nabla\bullet\mathbf{P}_f\right)\end{aligned} \tag{1}$$

In Eq.1 $e$, $m_f^*$, and $n_{0,f}$ are the electron's charge, effective mass, and density in the conduction band, $E_F$ is the Fermi energy, and $c$ is the speed of light in vacuum. For good conductors the ratio $E_F / m_f^* c^2$ is between $10^{-5}$ and $10^{-3}$, depending on effective masses, densities, and Fermi velocities. We choose $E_F / m_f^* c^2 = 10^{-4}$. The Lorentz force, $\dot{\mathbf{P}}_f \times \mathbf{H}$, is accompanied by a quadrupole-like Coulomb term proportional to $\mathbf{E}\left(\nabla \bullet \mathbf{P}_f\right)$, convective terms $\sim\left(\nabla \bullet \dot{\mathbf{P}}_f\right)\dot{\mathbf{P}}_f + \left(\dot{\mathbf{P}}_f \bullet \nabla\right)\dot{\mathbf{P}}_f$, and linear and nonlinear pressure terms proportional to $\nabla\left(\nabla \bullet \mathbf{P}_f\right)$ and $\left(\nabla \bullet \mathbf{P}_f\right)\nabla\left(\nabla \bullet \mathbf{P}_f\right)$, respectively. As we will see later, in nanocavity environments the linear pressure term can shift the band structures by tens of nanometers. If we assume that the fields and their respective polarizations and magnetizations may be decomposed as a superposition of harmonics, Eq.1 then represents a set of three coupled equations [28]. We distinguish between TE- and TM-polarized fields and define:

$$\mathbf{E} = \begin{pmatrix} E_x \\ E_y \\ E_z \end{pmatrix} = \begin{pmatrix} \mathbf{i}\left( E_{TEx}^{\omega} e^{-i\omega t} + \left(E_{TEx}^{\omega}\right)^* e^{i\omega t} + E_{TEx}^{2\omega} e^{-2i\omega t} + \left(E_{TEx}^{2\omega}\right)^* e^{2i\omega t} + E_{TEx}^{3\omega} e^{-3i\omega t} + \left(E_{TEx}^{3\omega}\right)^* e^{3i\omega t} \right) \\ +\mathbf{j}\left( E_{TMy}^{\omega} e^{-i\omega t} + \left(E_{TMy}^{\omega}\right)^* e^{i\omega t} + E_{TMy}^{2\omega} e^{-2i\omega t} + \left(E_{TMy}^{2\omega}\right)^* e^{2i\omega t} + E_{TMy}^{3\omega} e^{-3i\omega t} + \left(E_{TMy}^{3\omega}\right)^* e^{3i\omega t} \right) \\ +\mathbf{k}\left( E_{TMz}^{\omega} e^{-i\omega t} + \left(E_{TMz}^{\omega}\right)^* e^{i\omega t} + E_{TMz}^{2\omega} e^{-2i\omega t} + \left(E_{TMz}^{2\omega}\right)^* e^{2i\omega t} + E_{TMz}^{3\omega} e^{-3i\omega t} + \left(E_{TMz}^{3\omega}\right)^* e^{3i\omega t} \right) \end{pmatrix} \tag{2a}$$

$$\mathbf{H} = \begin{pmatrix} H_x \\ H_y \\ H_z \end{pmatrix} = \begin{pmatrix} \mathbf{i}\left( H_{TMx}^{\omega} e^{-i\omega t} + \left(H_{TMx}^{\omega}\right)^* e^{i\omega t} + H_{TMx}^{2\omega} e^{-2i\omega t} + \left(H_{TMx}^{2\omega}\right)^* e^{2i\omega t} + H_{TMx}^{3\omega} e^{-3i\omega t} + \left(H_{TMx}^{3\omega}\right)^* e^{3i\omega t} \right) \\ +\mathbf{j}\left( H_{TEy}^{\omega} e^{-i\omega t} + \left(H_{TEy}^{\omega}\right)^* e^{i\omega t} + H_{TEy}^{2\omega} e^{-2i\omega t} + \left(H_{TEy}^{2\omega}\right)^* e^{2i\omega t} + H_{TEy}^{3\omega} e^{-3i\omega t} + \left(H_{TEy}^{3\omega}\right)^* e^{3i\omega t} \right) \\ +\mathbf{k}\left( H_{TEz}^{\omega} e^{-i\omega t} + \left(H_{TEz}^{\omega}\right)^* e^{i\omega t} + H_{TEz}^{2\omega} e^{-2i\omega t} + \left(H_{TEz}^{2\omega}\right)^* e^{2i\omega t} + H_{TEz}^{3\omega} e^{-3i\omega t} + \left(H_{TEz}^{3\omega}\right)^* e^{3i\omega t} \right) \end{pmatrix} \tag{2b}$$

The polarization directions are noted in Fig. 1(a). For TE-polarization the H-fields point along the y- and z- directions; for TM-polarization the E-field has components along y and z. The oscillator model is exemplified by the following scaled equations that describe generic, rapidly varying field envelope functions valid in the pump depletion regime [28]:

$$\ddot{\mathbf{P}}_{b,\omega} + \tilde{\gamma}_{b,\omega}\dot{\mathbf{P}}_{b,\omega} + \tilde{\omega}_{0,b,\omega}^2 \mathbf{P}_{b,\omega} \approx \frac{n_{0,b} e^2 \lambda_0^2}{m_b^* c^2}\mathbf{E}_{\omega} + \frac{e\ \lambda_0}{m_b^* c^2}\begin{pmatrix} -\frac{1}{2}\mathbf{E}_{\omega}^*\nabla \bullet \mathbf{P}_{b,2\omega} \\ +2\mathbf{E}_{2\omega}\nabla \bullet \mathbf{P}_{b,\omega}^* \\ -\frac{2}{3}\mathbf{E}_{2\omega}^*\nabla \bullet \mathbf{P}_{b,3\omega} \\ -\frac{3}{2}\mathbf{E}_{3\omega}\nabla \bullet \mathbf{P}_{b,2\omega}^* \end{pmatrix} + \frac{e\ \lambda_0}{m_b^* c^2}\begin{pmatrix} \left(\dot{\mathbf{P}}_{b,\omega}^* + i\omega \mathbf{P}_{b,\omega}^*\right)\times \mathbf{H}_{2\omega} \\ +\left(\dot{\mathbf{P}}_{b,2\omega} - 2i\omega \mathbf{P}_{b,2\omega}\right)\times \mathbf{H}_{\omega}^* \\ +\left(\dot{\mathbf{P}}_{b,2\omega}^* + 2i\omega \mathbf{P}_{b,2\omega}^*\right)\times \mathbf{H}_{3\omega} \\ +\left(\dot{\mathbf{P}}_{b,3\omega} - 3i\omega \mathbf{P}_{b,3\omega}\right)\times \mathbf{H}_{2\omega}^* \end{pmatrix} \tag{3a}$$

$$\ddot{\mathbf{P}}_{b,2\omega} + \tilde{\gamma}_{b,2\omega}\dot{\mathbf{P}}_{b,2\omega} + \tilde{\omega}_{0,b,2\omega}^2 \mathbf{P}_{b,2\omega} \approx \frac{n_{0,b} e^2 \lambda_0^2}{m_b^* c^2}\mathbf{E}_{2\omega} + \frac{e\ \lambda_0}{m_b^* c^2}\begin{pmatrix} \mathbf{E}_{\omega}\nabla \bullet \mathbf{P}_{b,\omega} \\ -\frac{1}{3}\mathbf{E}_{\omega}^*\nabla \bullet \mathbf{P}_{b,3\omega} \\ -3\mathbf{E}_{3\omega}\nabla \bullet \mathbf{P}_{b,\omega}^* \end{pmatrix} + \frac{e\ \lambda_0}{m_b^* c^2}\begin{pmatrix} \left(\dot{\mathbf{P}}_{b,\omega} - i\omega \mathbf{P}_{b,\omega}\right)\times \mathbf{H}_{\omega} \\ +\left(\dot{\mathbf{P}}_{b,\omega}^* + i\omega \mathbf{P}_{b,\omega}^*\right)\times \mathbf{H}_{3\omega} \\ +\left(\dot{\mathbf{P}}_{b,3\omega} - 3i\omega \mathbf{P}_{b,3\omega}\right)\times \mathbf{H}_{\omega}^* \end{pmatrix} \tag{3b}$$

$$\ddot{\mathbf{P}}_{b,3\omega} + \tilde{\gamma}_{b,3\omega}\dot{\mathbf{P}}_{b,3\omega} + \tilde{\omega}_{0,b,3\omega}^2 \mathbf{P}_{b,3\omega} \approx \frac{n_{0,b} e^2 \lambda_0^2}{m_b^* c^2}\mathbf{E}_{3\omega} + \frac{e\ \lambda_0}{m_b^* c^2}\begin{pmatrix} \frac{1}{2}\mathbf{E}_{\omega}\nabla \bullet \mathbf{P}_{b,2\omega} \\ +2\mathbf{E}_{2\omega}\nabla \bullet \mathbf{P}_{b,\omega} \end{pmatrix} + \frac{e\ \lambda_0}{m_b^* c^2}\begin{pmatrix} \left(\dot{\mathbf{P}}_{b,2\omega} - 2i\omega \mathbf{P}_{b,2\omega}\right)\times \mathbf{H}_{\omega} \\ +\left(\dot{\mathbf{P}}_{b,\omega} - i\omega \mathbf{P}_{b,\omega}\right)\times \mathbf{H}_{2\omega} \end{pmatrix} \tag{3c}$$

The scaled coefficients are $\tilde{\gamma}_{b,N\omega} = (\gamma_b - Ni\omega)$, $\tilde{\omega}^2_{0,N\omega} = (\omega^2_{0,b} - (N\omega)^2 + i\gamma_b N\omega)$; N is an integer that denotes the given harmonic order; the subscript *b* stands for bound. For further details about the model and the integration scheme employed we direct the reader to reference [28].

Bulk second and third order nonlinearities may be introduced directly into each of Eqs.3, or by defining a nonlinear polarization in the usual way, i.e. $\mathbf{P}_{NL} = \chi^{(2)}\mathbf{E}^2 + \chi^{(3)}\mathbf{E}^3 + ...$, as we do. The second order polarization vector of GaAs may be written as follows [47]:

$$\begin{pmatrix} P^{(2)}_{NL,x} \\ P^{(2)}_{NL,y} \\ P^{(2)}_{NL,z} \end{pmatrix} = 2\mathrm{d}_{14} \begin{pmatrix} E_y E_z \\ E_x E_z \\ E_x E_y \end{pmatrix} \tag{4}$$

Substituting the E-field vector defined in Eq.2 into Eq.4 leads to the following equations:

$$P^{(2)}_{NL,x} = 2\mathrm{d}_{14} \begin{pmatrix} \left( E^{2\omega}_{TMy}\left(E^{\omega}_{TMz}\right)^* + \left(E^{\omega}_{TMy}\right)^* E^{2\omega}_{TMz} + E^{3\omega}_{TMy}\left(E^{2\omega}_{TMz}\right)^* + \left(E^{2\omega}_{TMy}\right)^* E^{3\omega}_{TMz} \right) e^{-i\omega t} \\ + \left( E^{\omega}_{TMy} E^{\omega}_{TMz} + E^{3\omega}_{TMy}\left(E^{\omega}_{TMz}\right)^* + \left(E^{\omega}_{TMy}\right)^* E^{3\omega}_{TMz} \right) e^{-2i\omega t} \\ + \left( E^{2\omega}_{TMy} E^{\omega}_{TMz} + E^{\omega}_{TMy} E^{2\omega}_{TMz} \right) e^{-3i\omega t} \end{pmatrix} \tag{5a}$$

$$P^{(2)}_{NL,y} = 2\mathrm{d}_{14} \begin{pmatrix} \left( E^{2\omega}_{TEx}\left(E^{\omega}_{TMz}\right)^* + \left(E^{\omega}_{TEx}\right)^* E^{2\omega}_{TMz} + E^{3\omega}_{TEx}\left(E^{2\omega}_{TMz}\right)^* + \left(E^{2\omega}_{TEx}\right)^* E^{3\omega}_{TMz} \right) e^{-i\omega t} \\ + \left( E^{\omega}_{TEx} E^{\omega}_{TMz} + E^{3\omega}_{TEx}\left(E^{\omega}_{TMz}\right)^* + \left(E^{\omega}_{TEx}\right)^* E^{3\omega}_{TMz} \right) e^{-2i\omega t} \\ + \left( E^{2\omega}_{TEx} E^{\omega}_{TMz} + E^{\omega}_{TEx} E^{2\omega}_{TMz} \right) e^{-3i\omega t} \end{pmatrix} \tag{5b}$$

$$P^{(2)}_{NL,z} = 2\mathrm{d}_{14} \begin{pmatrix} \left( E^{2\omega}_{TEx}\left(E^{\omega}_{TMy}\right)^* + \left(E^{\omega}_{TEx}\right)^* E^{2\omega}_{TMy} + E^{3\omega}_{TEx}\left(E^{2\omega}_{TMy}\right)^* + \left(E^{2\omega}_{TEx}\right)^* E^{3\omega}_{TMy} \right) e^{-i\omega t} \\ + \left( E^{\omega}_{TEx} E^{\omega}_{TMy} + E^{3\omega}_{TEx}\left(E^{\omega}_{TMy}\right)^* + \left(E^{\omega}_{TEx}\right)^* E^{3\omega}_{TMy} \right) e^{-2i\omega t} \\ + \left( E^{2\omega}_{TEx} E^{\omega}_{TMy} + E^{\omega}_{TEx} E^{2\omega}_{TMy} \right) e^{-3i\omega t} \end{pmatrix} \tag{5c}$$

A cursory inspection of Eqs.5 might lead one to prematurely conclude that if a TM-polarized field were incident on GaAs the only surviving source term would be a TE-polarized SH signal, namely $\mathbf{i}\, E^{\omega}_{TMy} E^{\omega}_{TMz} e^{-2i\omega t}$ in Eq.5a. Fortunately, the picture is far more interesting than it appears at first sight. Nonlinear source terms in both Eqs.1 and 3, i.e. free and bound charges in the metal and bound charges in GaAs (via derivatives of the type $\nabla \bullet \mathbf{P}_{b,N\omega}$ and Lorentz magnetic terms) give rise to TM-polarized harmonics. A more careful analysis of Eqs.5 then reveals that TM-polarized pump and harmonic fields serve as nonlinear sources for all TE-polarized harmonic fields, *including the pump*. The production of a TE-polarized pump field does not require the presence of the metal, and initiates with the mere introduction of the magnetic Lorentz force in Eqs.3. Once TE-polarized fields are generated, all interaction channels become active and the generation of all harmonic fields is assured. However, the generation of down-converted, orthogonally polarized pump photons depends on the structure of the $\chi^{(2)}$ tensor. For example, a $\chi^{(2)}$ tensor whose only non-zero components are $d_{11}$, $d_{22}$, and $d_{33}$ cannot couple TM- to TE-polarized pump photons. We will return to this issue below.

The description of $\chi^{(3)}$ contributions may begin with the general expansion of the third order polarization as follows [47]:

$$P_{NL,i}^{(3)} = \sum_{j=1,3} \sum_{k=1,3} \sum_{l=1,3} \chi_{i,j,k,l} E_j E_k E_l \tag{6}$$

For a material like GaAs having cubic symmetry of the type $\bar{4}3m$, Eq.6 reduces to:

$$\begin{aligned} P_{NL,x}^{(3)} &= \chi_{xxxx}^{(3)} E_x^3 + 3\chi_{xxyy}^{(3)} E_y^2 E_x + 3\chi_{xxzz}^{(3)} E_z^2 E_x \\ P_{NL,y}^{(3)} &= \chi_{yyyy}^{(3)} E_y^3 + 3\chi_{xxyy}^{(3)} E_x^2 E_y + 3\chi_{yyzz}^{(3)} E_z^2 E_y \\ P_{NL,z}^{(3)} &= \chi_{zzzz}^{(3)} E_z^3 + 3\chi_{zzxx}^{(3)} E_x^2 E_z + 3\chi_{zzyy}^{(3)} E_y^2 E_z \end{aligned} \tag{7}$$

For the metal the situation is similar, except that for isotropic crystal symmetry the relations between the tensor components allow one to write:

$$\begin{aligned} P_{NL,x}^{(3)} &= \chi_{Ag}^{(3)} \left( E_x^3 + E_y^2 E_x + E_z^2 E_x \right) \\ P_{NL,y}^{(3)} &= \chi_{Ag}^{(3)} \left( E_y^3 + E_x^2 E_y + E_z^2 E_y \right) \\ P_{NL,z}^{(3)} &= \chi_{Ag}^{(3)} \left( E_z^3 + E_x^2 E_z + E_y^2 E_z \right) \end{aligned} \tag{8}$$

It is evident that substitution of Eqs.2 for the electric field into Eqs.7-8 leads to $\chi^{(3)}$ contributions to all harmonic components, with self- and cross-phase modulation of the fields along with terms that couple orthogonal polarization states. The introduction of phase modulation effects on the pump fields is important, especially in metals [11, 12], because given the right combination of intense fields and large $\chi^{(3)}$ band shifts may be substantial.

## 4. Nonlinear results for Silver gratings filled with GaAs

Whether it is due to vertical or horizontal resonances or a combination of both, EOT at near-IR, visible and UV wavelengths is characterized by field localization, absorption, and penetration inside the metal because in these ranges metals display dielectric constants of order unity. The interaction of light with free and bound electrons in metals becomes more efficient especially if the light is concentrated in small volumes. We consider the same system described in Figs.1 and 2. Incident pulses are 120fs with peak intensities ~2GW/cm$^2$. Pulse duration is such that the FP resonance is nearly resolved. To avoid confusion we temporarily introduce bulk quadratic and cubic nonlinearities only in GaAs. Later we will relax this condition. For illustration purposes we choose $\chi^{(2)}=2d_{14}=2d_{25}=2d_{36}=10$pm/V, and $\chi^{(3)}_{xxxx}=\chi^{(3)}_{yyyy}=\chi^{(3)}_{zzzz}=3\chi^{(3)}_{xxyy}=3\chi^{(3)}_{xxzz}=3\chi^{(3)}_{yyzz}=3\chi^{(3)}_{zzyy}=3\chi^{(3)}_{zzxx}=3\chi^{(3)}_{xxyy}=10^{-19}\,(\mathrm{m}^2/\mathrm{V}^2)$. An incident TM-polarized field generates at least four harmonic fields: TM-polarized SH and TH (Fig. 3(a-b)), and TE-polarized SH and TH (Fig. 4(a-b)). If these results are examined together with Fig. 2(a) they reveal that the nonlinear response is influenced by the linear properties: all the generated harmonics experience the same forbidden states of the incident pump field. This situation occurs because the harmonics operate at twice and three times the frequency of the pump, but experience the same index of refraction as the pump. Once the fields are generated and leave the slit's proximity, where coupling is largest, the harmonics are influenced and constrained by the size of the wavelength relative to the grating periodicity. For example, the TM-polarized SH is inhibited beginning when *p* matches the unperturbed air/silver plasmon wavelength of the pump and SH. The same phenomenon is evident also for THG with appropriate pitch values.

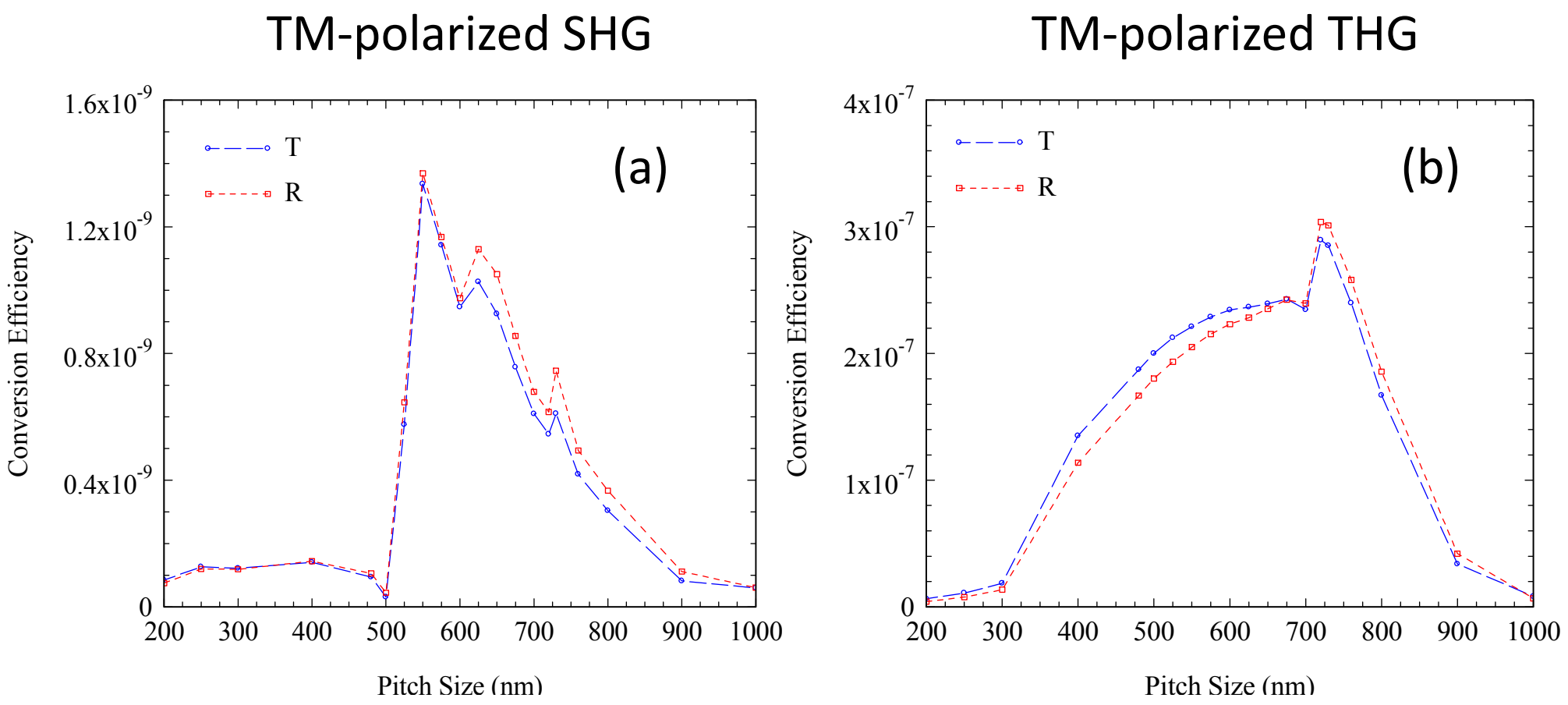

Fig.3. TM-polarized (a) SH and (b) TH transmitted (blue line–circle markers), reflected (red line–square markers) conversion efficiencies. SH energies are comparable for p>500nm and p<500nm. However, for p>500nm SH light immediately leaves the grating; for p<500nm the fields linger near the grating and are re-absorbed by the metal.

Although the choice $\chi^{(2)}=10$pm/V in GaAs leads to predicted TE- and TM-polarized SH efficiencies having similar values, efficiencies for TM-polarized SHG [20] have not been reported. The efficiency that we predict for the SH TM-polarized signal (which is independent of GaAs and arises mostly from the metal) is nearly 100 times larger than SHG from smooth metal layers, and 10 times larger than SHG from metal-dielectric stacks [28]. The lack of plasmonic resonances at the SH and TH wavelengths (see Fig. 2(b)) suggests that this behavior depends almost exclusively on (FP) cavity-Q [41], field overlap, and phase-locking.

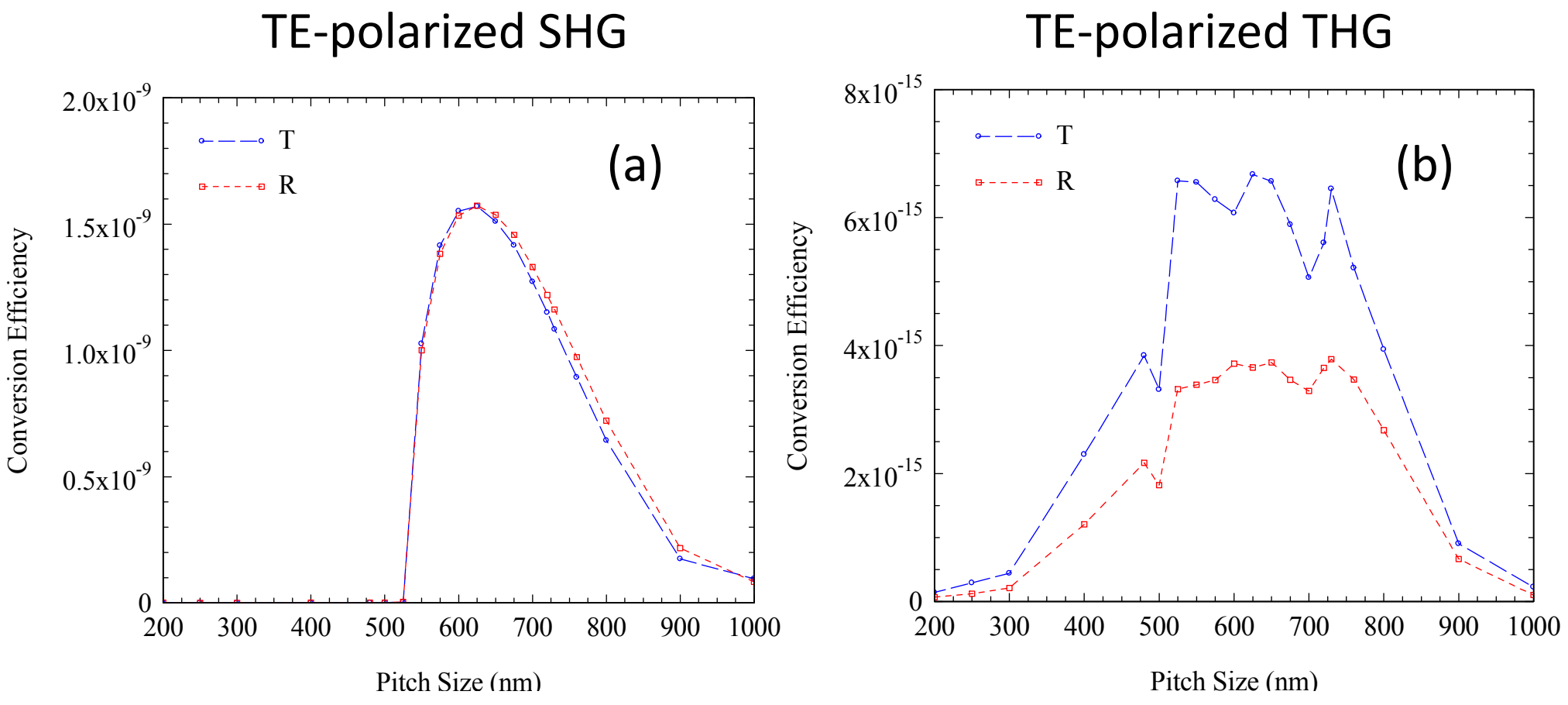

Fig.4: TE-polarized (a) second and (b) third harmonic transmitted (blue line–circle markers), reflected (red line–square markers) total conversion efficiencies. The account of the dynamics that we provided in Figs.3 for TM-polarized SH fields also applies to the TE-polarized SH signal.

## 5. Spectral features and field profiles

We now examine the individual features of the harmonic signals emitted by the grating, including spectra and field profiles. We choose pump pulses having peak intensity 2GW/cm$^2$ and 50fs duration, incident on a grating with *p*=590nm. This pitch optimizes SHG of both polarizations and places plasmonic features away from all wavelengths. For GaAs we choose $\chi^{(2)}$=100pm/V and $\chi^{(3)}=10^{-18}$ $m^2/V^2$, consistent with more recent experimental observations of harmonic generation in similar wavelength ranges [37, 38]. In Figs.5-6 we show reflected and transmitted spectra for all generated fields. We note that the spectra display a small shift between transmission and reflection maxima but have similar amplitudes. In Fig.6 we show the spectra for TE-polarized pump photons. The conversion efficiencies that we predict for this novel process are already comparable to TE-polarized, TH conversion efficiencies. Albeit relatively small, the efficiency of this down-conversion process may be enhanced in bulk GaAs or by pumping the grating with TM-polarized SH and TH seed fields, as Eq.5a suggests, via the term: $\mathbf{i}\left(E_{TMy}^{3\omega}\left(E_{TMz}^{2\omega}\right)^{*}+\left(E_{TMy}^{2\omega}\right)^{*}E_{TMz}^{3\omega}\right)e^{-i\omega t}$.

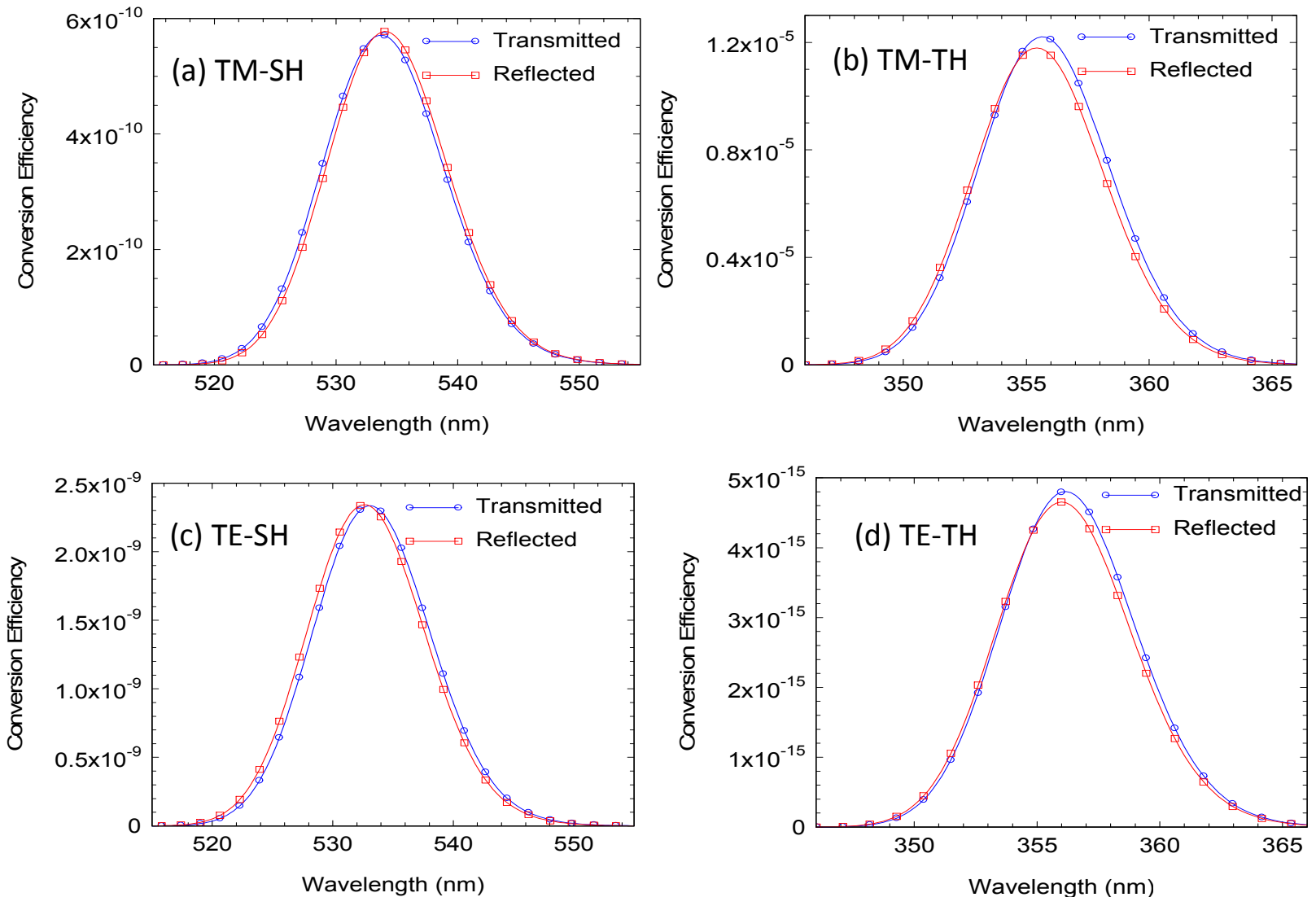

Fig.5. TM-polarized SH (a) and TH (b) transmitted and reflected conversion efficiency spectra, normalized with respect to the spectrum of the transmitted pump field. (c) and (d): same as (a) and (b) but for TE-polarized fields. Predicted conversion efficiency of the TM-polarized TH signal is remarkably high.

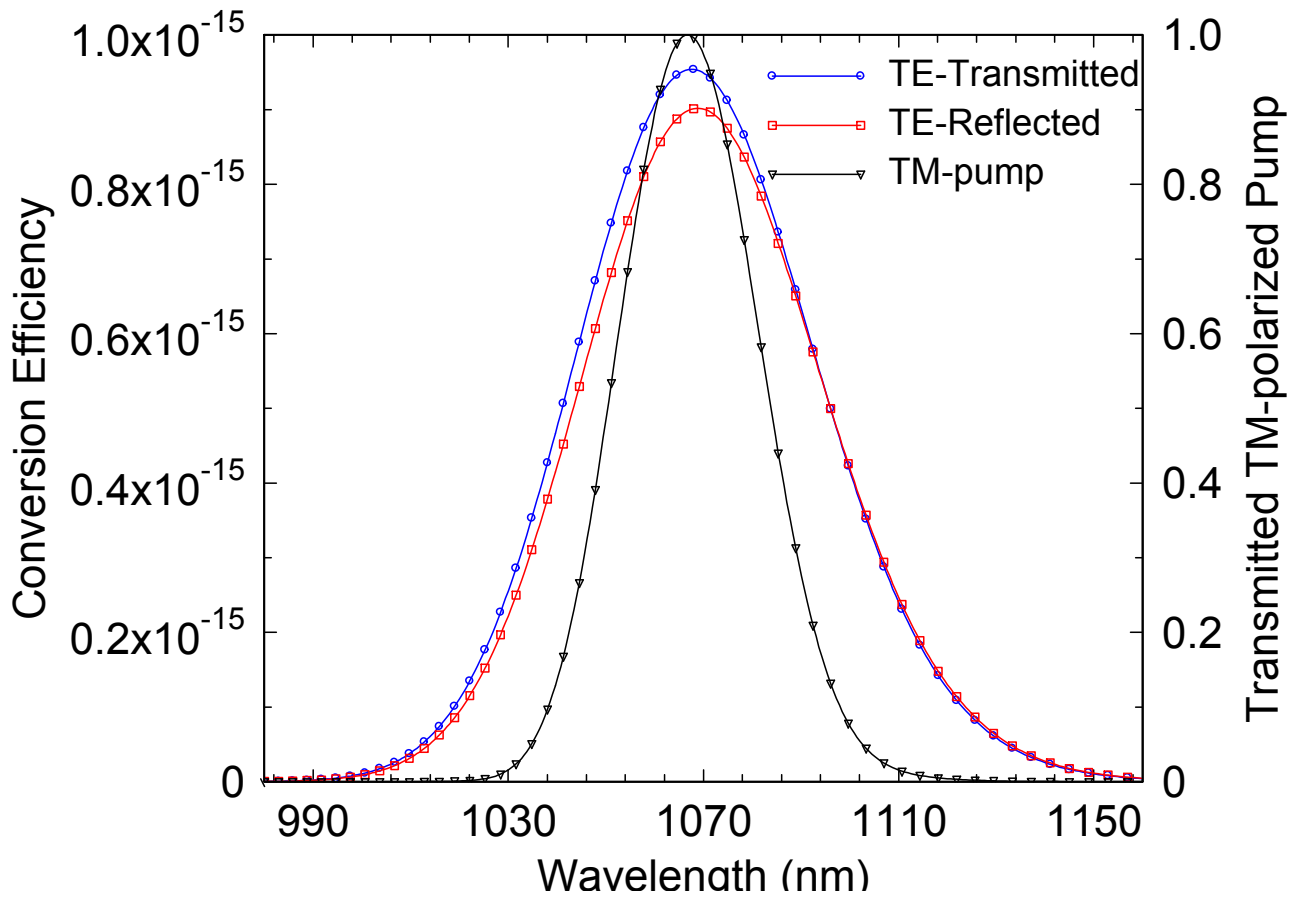

Fig.6. Spectra of the transmitted, TM-polarized pump field (black line – triangle markers), and TE-polarized, transmitted (blue line – circle markers) and reflected (red line – square markers) down-converted pump photons. Conversion efficiencies are relatively small, but are nevertheless similar to those of TE-polarized THG.

In Fig.7 we show snapshots of the corresponding field profiles inside the nanocavity when the peak of the pulse reaches the grating. All fields are well-localized inside the cavity, including the TH tuned at 354nm. Similar resonant behavior was obtained for planar cavities with harmonic fields tuned below the absorption edge [38, 48]. One of our objectives is to also show that phase-locking [29-35] is playing a non trivial role. A smooth, 100nm-thick GaAs layer is only 20% transparent at 532nm, and completely opaque at 354nm. In a multi-pass geometry or a resonant cavity environment [48] the homogenous portion of the SH signal is removed more efficiently compared to bulk, so that all generated components that survive in the nonlinear medium propagate under phase-locking conditions. In our case –Fig. 2(b)– linear transmission is less than 0.05% at 532nm. At 354nm the grating is a bit

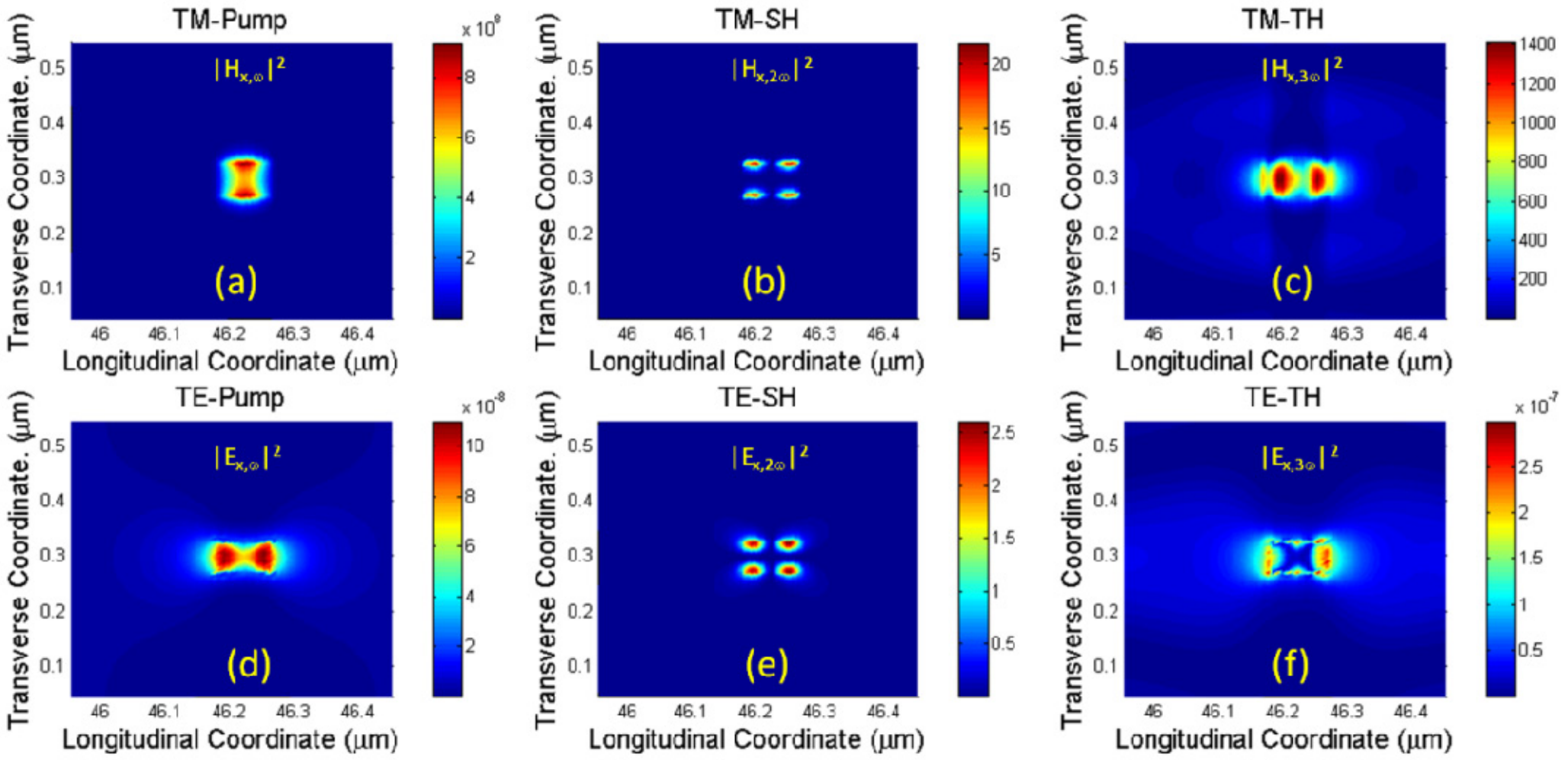

Fig.7. Pump and harmonic field intensities inside and near the nano-cavities. The magnetic field intensities are depicted for TM-polarization; the transverse electric field is shown for TE-polarization. The pump magnetic field intensity (a) is amplified 250 times; the transverse pump electric field (not shown) is amplified by approximately two orders of magnitude. This combination gives way to TH conversion efficiencies that are unusually large ($\sim 10^{-5}$).

transmissive thanks to the natural transparency of silver, but GaAs remains completely opaque.

More convincing numerical evidence of phase-locking may be achieved by increasing substrate thickness up to 170nm, and by reducing the width of the slits down to 20nm, so that we are still operating under resonant conditions (see Fig. 1(b)). As a result conversion efficiencies and localization properties vary little compared to Figs.6-7, except for an increased number of longitudinal peaks, as the fields resonate inside the cavity even though all the TE-generated fields are far below cut-off. This is a sure sign that phase-locking is the mechanism that drives the harmonic fields to resonate, despite the fact that the cavity should resonate only at the pump frequency [38, 48].

## 6. Nonlinear results for Silver grating filled with a diagonal nonlinear material

We now show that it is possible to improve SH and TH conversion efficiencies if we assume the medium has a $\chi^{(2)}$ tensor whose only non-zero components are $d_{11}=d_{22}=d_{33}$, and if a $\chi^{(3)}$ is triggered inside the metal. The second order polarization of Eq.4 takes the following form:

$$\begin{pmatrix} P_{NL,x}^{(2)} \\ P_{NL,y}^{(2)} \\ P_{NL,z}^{(2)} \end{pmatrix} = d_{11} \begin{pmatrix} E_x^2 \\ E_y^2 \\ E_z^2 \end{pmatrix} . \qquad (9)$$

Unlike Eqs.5, Eq.9 allows full exploitation of transverse and longitudinal field localization, since in a cavity environment orthogonally polarized fields may not have similar amplitudes and localization properties. The third order nonlinear coefficient in metal sections is chosen as in Eq.8. For our calculations we use $\chi^{(3)}_{Ag} \sim 100\chi^{(3)}_{GaAs}$, where $\chi^{(3)}_{GaAs} = 10^{-18} m^2/V^2$, as before. Our choice corresponds to $\chi^{(3)}_{Ag} \sim 0.75\times10^{-8}$ esu, which is much smaller than what was reported for Cu ( $\chi^{(3)}_{Cu} \sim 10^{-6}$ esu) [48-51] in a multilayer environment [11], but is in line with nonlinearities reported for silver particles. For illustration purposes we fill the cavity with a material that has the same index of refraction as GaAs, the $\chi^{(2)}$ of Eq.9, and we compare efficiencies with all else being equal. In Figs.8 we depict the conversion efficiencies for TM-polarized harmonics (TE-polarized fields remain null) as a function of pulse duration. The figures show that efficiencies increase as a function of pulse width and saturate for pulses longer than ~60fs. This behavior is typical of cavity phenomena [48]. The most important features in Figs.8 are perhaps the total efficiencies, i.e. ~$10^{-6}$ for SHG and ~$10^{-3}$ for THG. By choosing a suitable material and by allowing a non-zero $\chi^{(3)}$ in the metal, results in exceptional conversion efficiencies in wavelength ranges that are usually deemed inaccessible, independent of phase matching conditions and absorption at the harmonic wavelengths.

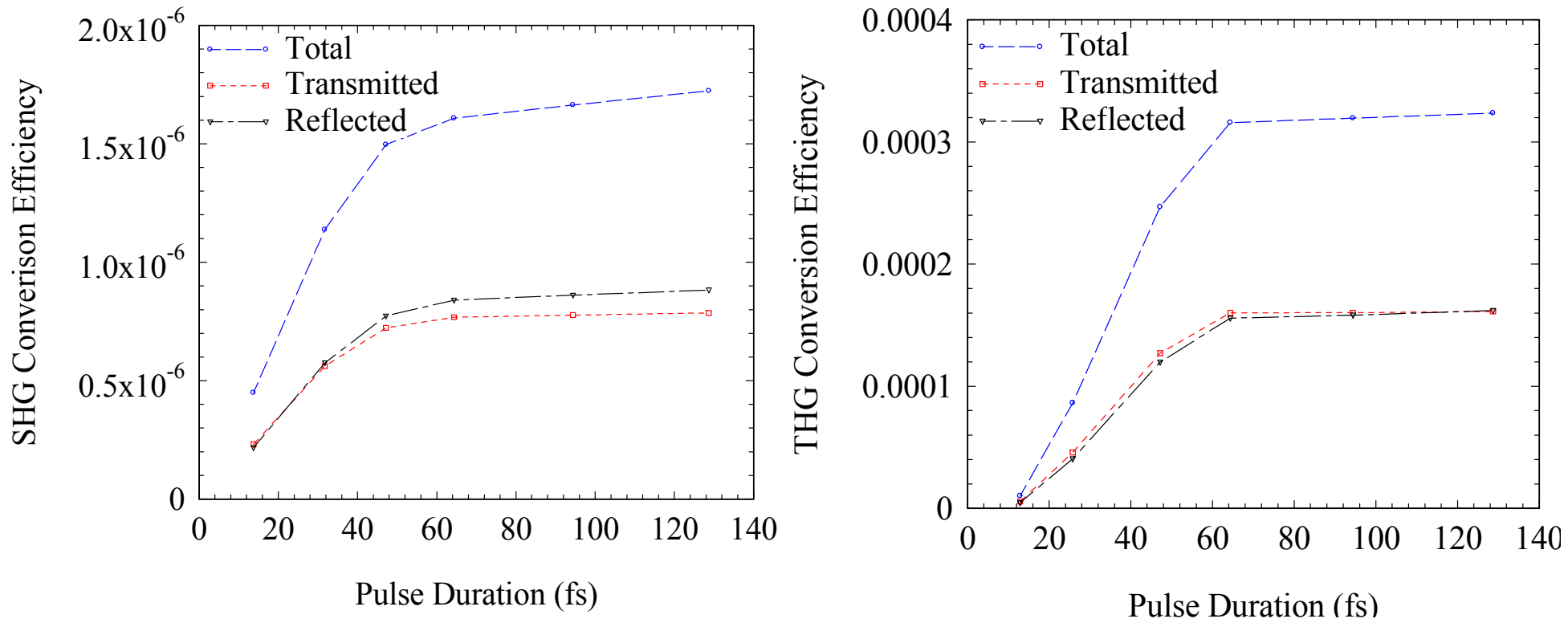

Fig.8. SH (a) and TH (b) conversion efficiencies vs. pulse duration. A diagonal $\chi^{(2)}$ tensor boosts SHG by at least three orders of magnitude compared to GaAs thanks to the full exploitation of field localization properties. Allowing a non-zero $\chi^{(3)}$ in the metal improves THG well in excess of one order of magnitude compared to GaAs alone.

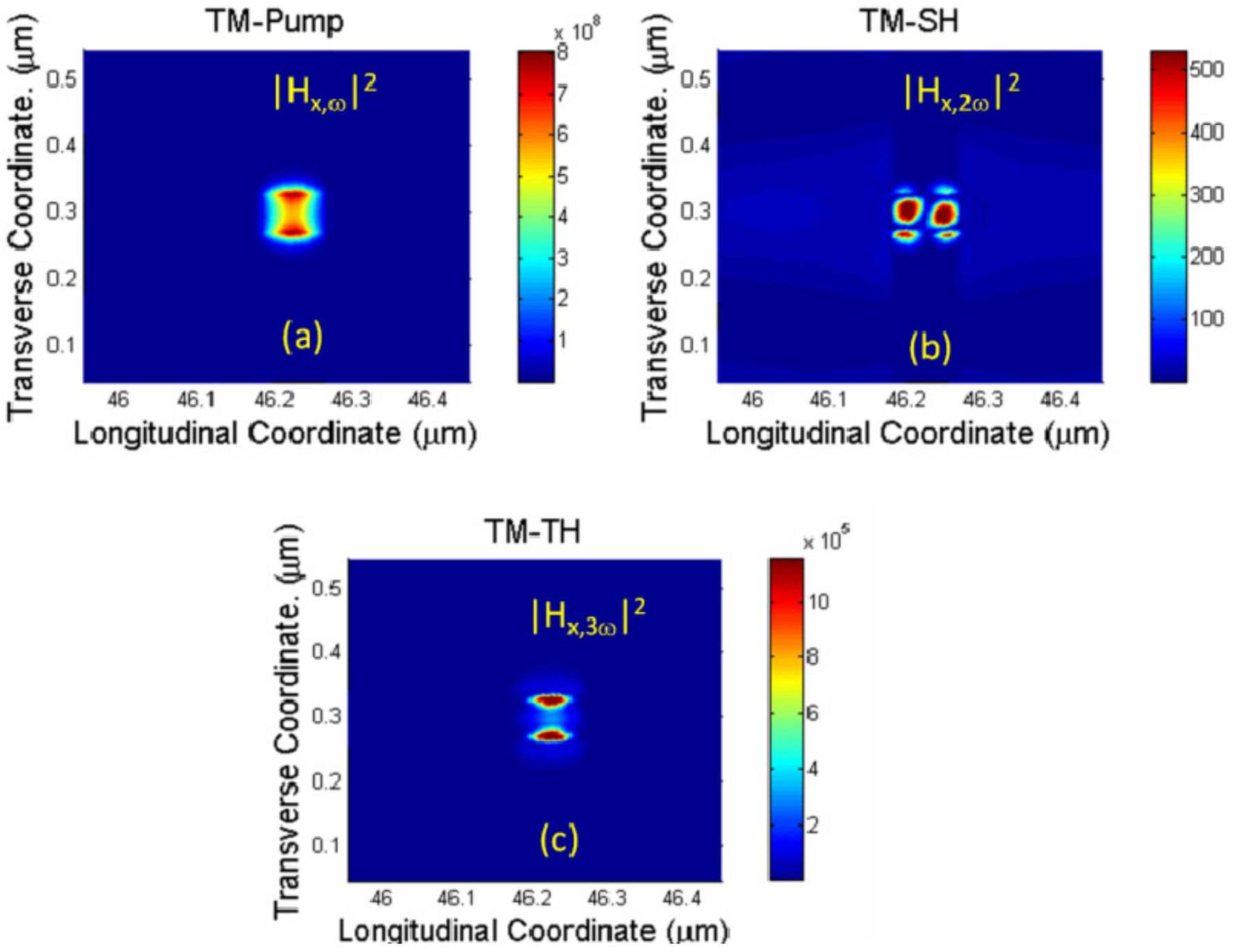

Fig.9: Typical pump (a), SH (b) and TH (c) magnetic field intensities when the peak of a 50fs pulse reaches the grating. One should compare the field with the corresponding harmonics in Fig.7. While the entire metal surface has a non-zero $\chi^{(3)}$, only nonlinear sources on the metal walls inside the cavity matter to the process.

Finally in Fig.9 we show typical field profiles that correspond to Fig.8. The figure should be compared directly with the corresponding intensities of Fig.7. While in Fig.7 most of the TM-polarized SH-signal came mostly from the metal, now the SH field originates mostly from the bulk nonlinearities of GaAs. Nevertheless, in Fig. 9(b) one can still see remnants of the localization displayed in Fig. 7(b), since the metal remains an active participant. In contrast, THG originates at the walls of the slit: the field spills into the cavity, becomes phase-locked and resonates with the pump, leading to large conversion efficiencies in the UV range.

**7. Electron gas pressure**

The discussion above suggests that in sub-wavelength regions and where metallic edges or corners are present the linear dielectric response is modulated by the pressure term. We analyze the linear regime of Eq.1, which may be written as follows:

$$\ddot{\mathbf{P}}_f + \tilde{\gamma}\dot{\mathbf{P}}_f = \frac{n_{0f}e^2}{m_f^*}\left(\frac{\lambda_0}{c}\right)^2 \mathbf{E} + \frac{5}{3}\frac{E_F}{m_f^* c^2}\nabla\left(\nabla \bullet \mathbf{P}_f\right) \quad . \tag{10}$$

Fourier transformation of Eq.10 leads to:

$$\mathbf{P}_f(\omega,k) = \alpha\mathbf{E}(\omega,k) + \beta\mathbf{K}\left[\mathbf{K}\bullet\mathbf{P}_f(\omega,k)\right] \quad , \tag{11}$$

where $\alpha = -\frac{n_{0,f}e^2}{\left(\omega^2 + i\tilde{\gamma}\omega\right)m_f^*}\left(\frac{\lambda_0}{c}\right)^2$, $\beta = \frac{1}{\left(\omega^2 + i\tilde{\gamma}\omega\right)}\frac{5}{3}\frac{E_F}{m_f^* c^2}$, and $\mathbf{K} = \mathbf{j}k_y + \mathbf{k}\,k_z$. Contributions from evanescent wave vectors may be determined by writing the linear solutions for the polarizations in terms of the longitudinal and transverse electric fields as follows:

$$P_{f,y}=\alpha\left(\frac{\left[\left(1-\beta\left(k_y^2+k_z^2\right)+\beta^2k_y^2k_z^2\right)\right]E_y+\beta k_yk_z\left[\left(1-\beta k_y^2\right)E_z\right]}{\left(1-\beta k_y^2\right)\left(1-\beta\left(k_y^2+k_z^2\right)\right)}\right)$$
$$P_{f,z}=\alpha\left[\frac{\left(1-\beta k_y^2\right)E_z+\beta k_zk_yE_y}{\left(1-\beta\left(k_y^2+k_z^2\right)\right)}\right] \quad (12)$$

The poles are removed by the complex nature of $\alpha$ and $\beta$ and the polarization is subject to resonant behavior. In Fig.10 we show the shifts that occur by varying the pressure coupling coefficient for the array of Fig.1. These shifts are substantial by any measure, even though cavity-Q is in the hundreds. The process can impact the entire band structure of the grating, with shifts and possible reshaping of linear and nonlinear spectra, suggesting that in these environments electron gas pressure should probably always be assessed.

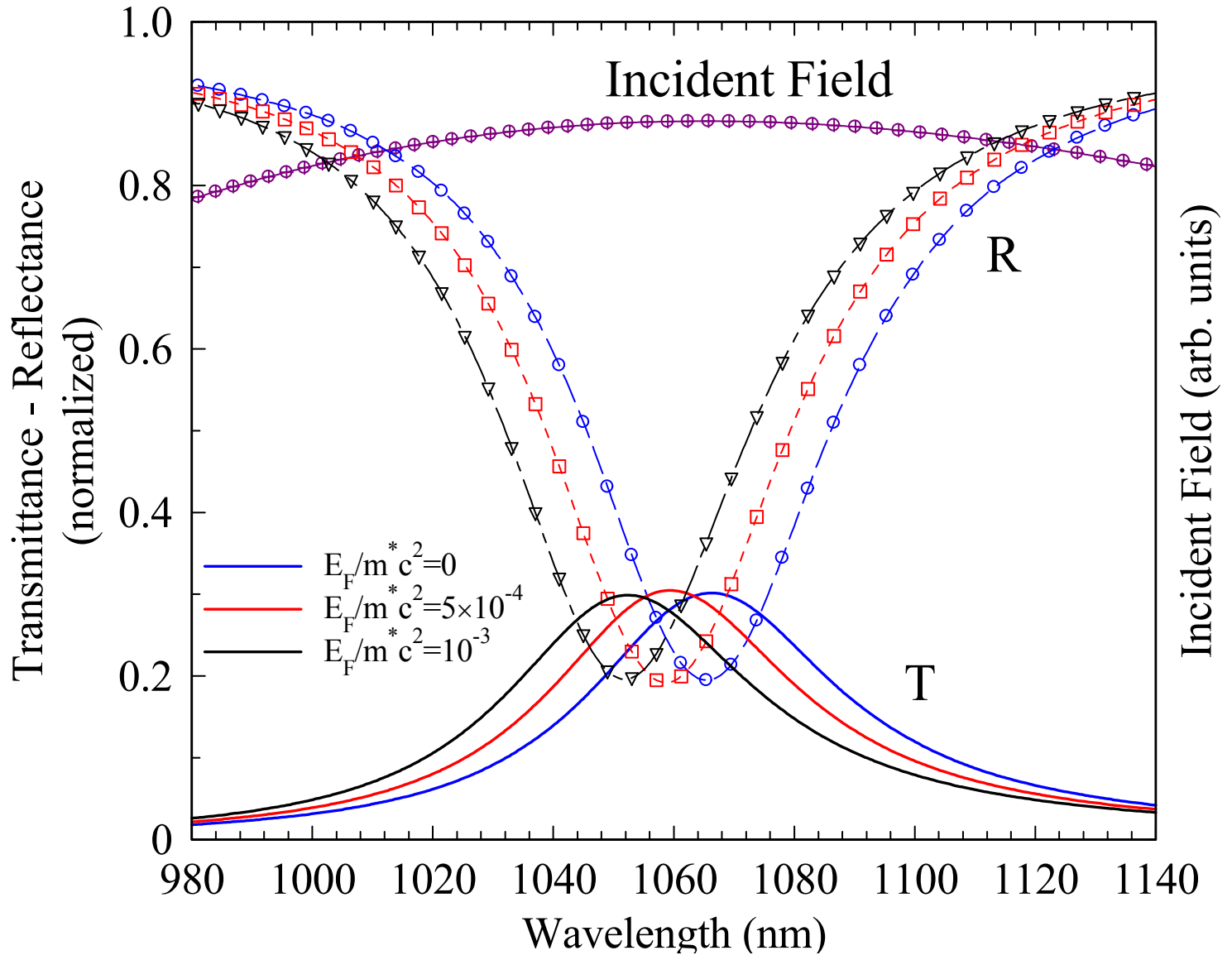

Fig.10. Transmission (solid lines) and reflection (dashed lines) vs. wavelength for the GaAs-filled array of Fig.1, for different values of $E_F/m^*c^2$. The incident field spectrum (purple, circled line) is also shown. Shifts due to $\chi^{(3)}$ in either GaAs and/or the metal may be of the same order of magnitude and could counteract the shifts depicted in this figure.

## 8. Conclusions

We have performed a study of second and third harmonic generation in semiconductor- and dielectric- filled silver gratings, in the EOT regime. We find that using a 2GW/cm$^2$ pump it is possible to achieve conversion efficiencies of $10^{-6}$ and $10^{-3}$ for SHG and THG, respectively, in regions dominated by FP cavity resonances at the pump wavelength, and by absorption at the harmonic wavelengths. A phase-locking mechanism binds the harmonics to the pump field and creates conditions that allow SH and TH fields to resonate despite the large nominal absorption. Plasmonic phenomena are unimportant to the process, and do not seem to offer an alternative to the sort of efficiencies that we predict by exploiting field enhancements in more traditional cavity environments. We have analyzed the interaction in two dimensions, by including surface and volume nonlinear phenomena in both metal and semiconductor sections of the grating, by considering different types of $\chi^{(2)}$ tensors, and by allowing the metal to display a $\chi^{(3)}$ response. We have shown that it is possible to trigger a novel down-conversion process that can re-generate phase-locked pump photons of polarization orthogonal compared to the incident pump field, i.e. a nonlinear polarizer of sorts. $\chi^{(3)}$ contributions and electron pressure within the metal can play relevant roles by boosting and/or shifting linear and nonlinear spectral features of the array. Further improvements to the individual conversion efficiencies are possible, especially for THG, by using metals like copper. If indeed $\chi^{(3)}_{Cu} \sim 10^{-6}$, then a few tens of MW/cm$^2$ may suffice to begin to deplete the pump, within the limits imposed by inevitable band shifts arising from self- and cross-phase modulation. Both qualitative and quantitative aspects of the nonlinear response are promising, considering that typical material thickness is far smaller than the coherence length of the nonlinear crystal, and confined to deeply sub-wavelength slits.

**9. References**

1. T.W. Ebessen, H. J. Lezec, H. F. Ghaemi, T. Thio, and P. A. Wolff, “Extraordinary optical transmission through subwavelength hole arrays”, Nature **391**, 667-669 (1998).
2. L. Salomon, F. Grillot, A. V. Zayats, and F. de Fornel, “Near-field distributionof optical transmission through sub-wavelength hole arrays,” *Phys. Rev. Lett.* **86**, 1110 (2001).
3. Y. Liu and S. Blair, “Fluorescence enhancement from array of sub-wavelength metal apertures,” *Opt. Lett.* **28**, 507 (2003).
4. D. J. Park, S. B. Choi, Y. H. Ahn, F. Rotermund, I. B. Sohn, C. Kang, M. S. Jeong, and D. S. Kim, “Terahertz near-field enhancement in narrow rectangular apertures on metal film,” *Opt. Express* **17**, 12493 (2009).
5. A. Nahata, R. A. Linke, T. Ishi, and K. Ohashi, “Enhanced nonlinear optical conversion using periodic nanostructured metal films,” *Opt. Lett.* **28**, 423 (2003).
6. M. Airola, Y. Liu, and S. Blair, "Second-harmonic generation from an array of sub-wavelength metal apertures, "*J. Opt. A: Pure Appl. Opt.* **7**, S118 (2005).

7. A. Lesuffler, L. Kiran Swaroop Kumar, and R. Gordon, "Enhanced second harmonic generation from Nanoscale double-hole arrays in gold film," *Appl. Phys. Lett.* **88**, 261104 (2006).
8. J. A. H. van Nieuwstadt, M. Sandtke, R. H. Harmsen, F. B. Segerink, J. C. Prangsma, S. Enoch, and L. Kuipers, "Strong Modification of the Nonlinear Optical Response of Metallic Subwavelength Hole Arrays," Phys. Rev. Lett. **97**, 146102 (2006).
9. N. Rakov, F. E. Ramos and M. Xiao, "Strong second harmonic generation from a thin silver film with randomly distributed small holes," *J. Phys.: Cond. Matter* **15**, L349 (2003).
10. T. Xu, X. Jiao and S. Blair, "Third-harmonic generation from arrays of subwavelength metal apertures," Opt. Express **17**, 23582 (2009).
11. Nick N. Lepeshkin, Aaron Schweinsberg, Giovanni Piredda, Ryan S. Bennink, and Robert W. Boyd, "Enhanced Nonlinear Optical Response of One-Dimensional Metal-Dielectric Photonic Crystals," Phys. Rev. Lett. **93**, 123902 (2004).
12. D. T. Owens, C. Fuentes-Hernandez, J. M. Hales, J. W. Perry, and B. Kippelen, "A comprehensive analysis of the contributions to the nonlinear optical properties of thin Ag films," J. Appl. Phys. **107**, 123114 (2010).
13. D. Krause, C. W. Teplin, and C. T. Rogers, "Optical surface second harmonic measurements of isotropic thin-film metals: Gold, silver, copper, aluminum, and tantalum, "J. Appl. Phys. **96**, 3626 (2004).
14. F. Xiang Wang, F. J. Rodríguez, W. M. Albers, R. Ahorinta, J. E. Sipe, and M. Kauranen, "Surface and bulk contributions to the second-order nonlinear optical response of a gold film," Phys. Rev. B **80**, 233402 (2009).
15. Y. R. Shen, *The Principles of Nonlinear Optics* ,Wiley Classics Library, Wiley, New York, (2002).
16. D. Maystre, M. Neviere, and R. Reinish, "Nonlinear polarization inside metals: a mathematical study of the free electron model," Appl. Phys. A **39**, 115 (1986).
17. M.A. Vincenti, V. Petruzzelli, A. D'Orazio, F. Prudenzano, M.J. Bloemer, N. Aközbek, and M. Scalora, "Second harmonic generation from nanoslits in metal substrates: applications to palladium-based $H_2$ sensor," *J. Nanophoton.* **2**, 021851 (2008).

18. F. I. Baida and D. Van Labeke, “Light transmission by subwavelength annular aperture arrays in metallic films,” *Optics Communications* **209**, 17 (2002).
19. F.I. Baida, D. Van Labeke, G. Granet, A. Moreau and A. Belkhir, “Origin of the super-enhanced light transmission through a 2-D metallic annular aperture array: a study of photonic bands,” *App Phys B* **79**, 1 (2004).
20. W. Fan, S. Zhang, N. C. Panoiu, A. Abdenour, S. Krishna, R. M. Osgood, K. J. Malloy, and S. R. J. Brueck, “Second Harmonic generation from a nanopatterned isotropic nonlinear material,” *Nano Letters* **6**, 1027 (2006).
21. E. H. Barakat, M. P. Bernal and F. I. Baida, “Second harmonic generation enhancement by use of annular aperture arrays embedded into silver and filled with lithium niobate,” *Opt. Express* **18**, 6530 (2010).
22. W. Fan, S. Zhang, K. J. S. Malloy, S. R. J. Brueck, N.C. Panoiu, and R. M. Osgood, “Second Harmonic generation from patterned GaAs inside a subwavelength metallic hole array,” *Opt. Express* **14**, 9570 (2006).
23. N. Bloembergen, R. K. Chang, S. S. Jha, C. H. Lee, “Optical harmonic generation in reflection from media with inversion symmetry,” Phys. Rev. **174** (3), 813 (1968).
24. J. E. Sipe, V. C. Y. So, M. Fukui and G. I. Stegeman, “Analysis of second-harmonic generation at metal surfaces,” Phys. Rev. B **21**, 4389 (1980).
25. J. E. Sipe and G. I. Stegeman, *Surface Polaritons: Electromagnetic Waves at Surfaces and Interfaces*, V. M. Agranovich and D. Mills North-Holland, Amsterdam (1982).
26. M. Corvi and W. L. Schaich, "Hydrodynamics model calculation of second harmonic generation at a metal surface," Phys. Rev. B **33**, 3688 (1986).
27. A. Eguiluz and J. J. Quinn, "Hydrodynamic model for surface plasmon in metals and degenerate semiconductors," Phys. Rev. B **14**, 1347 (1976).
28. M. Scalora, M. A. Vincenti, D. de Ceglia, V. Roppo, M. Centini, N. Akozbek, and M. J. Bloemer, “Second and Third Harmonic Generation in Metal-Based Nanostructures,” *Phys. Rev. A* **82**, 043828 (2010).
29. N. Bloembergen, and P. S. Pershan, “Light Waves at the Boundary of Nonlinear Media,” *Phys. Rev.* **128**, 606 (1962).
30. W. Glenn, “Second-harmonic generation by picosecond optical pulses,” IEEE J. Quantum Electron. **5**, 284-290 (1969).

31. J. T. Manassah and O. R. Cockings, "Induced phase modulation of a generated second-harmonic signal," Opt. Lett. **12**, 12 (1987).
32. S. L. Shapiro, "Second harmonic generation in LiNbO3 by picosecond pulses,"Appl. Phys. Lett. **13**, 19 (1968).
33. L. D. Noordam, H. J. Bakker, M. P. de Boer, and H. B. van Linden van den Heuvell, "Second-harmonic generation of femtosecond pulses: observation of phase-mismatch effects,"Opt. Lett. **15**, 24 (1990).
34. N.C. Khotari and X. Carlotti, Transient second-harmonic generation: influence of effective group-velocity dispersion," J. Opt. Soc. Am. B **5**, 756 (1988).
35. V. Roppo, M. Centini, C. Sibilia, M. Bertolotti, D. de Ceglia, M. Scalora, N. Akozbek, M. J. Bloemer, J. W. Haus, O. G. Kosareva and V. P. Kandidov, "Role of phase matching in pulsed second-harmonic generation: Walk-off and phase-locked twin pulses in negative-index media," Phys. Rev. A **76**, 033829 (2007).
36. V. Roppo, M. Centini, D. de Ceglia, M. A. Vincenti, J. W. Haus, N. Akozbek, M. J. Bloemer, and M. Scalora, "Anomalous momentum states, non-specular reflections, and negative refraction of phase-locked, second-harmonic pulses," Metamaterials **2**, 135 (2008).
37. M. Centini, V. Roppo, E. Fazio, F. Pettazzi, C. Sibilia, J. W. Haus, J. V. Foreman, N. Akozbek, M. J. Bloemer, and M. Scalora, "Inhibition of Linear Absorption in Opaque Materials Using Phase-Locked Harmonic Generation," Phys. Rev. Lett. **101**, 113905 (2008).
38. V. Roppo, C. Cojocaru, F. Raineri, G. D'Aguanno, J. Trull, Y. Halioua, R. Raj, I. Sagnes, R. Vilaseca, and M. Scalora, "Field localization and enhancement of phase-locked second- and third-order harmonic generation in absorbing semiconductor cavities" , Phys. Rev. A **80**, 043834 (2009).
39. V. Roppo, J. Foreman, N. Akozbek, M.A. Vincenti, and M. Scalora, "Third Harmonic Generation at 223nm in the Metallic Regime of GaP," http://arxiv.org/abs/1011.0627 (2010).
40. E.D. Palik, *Handbook of Optical Constants of Solids* (Academic Press, London-New York (1985).

41. M. A. Vincenti, M. De Sario, V. Petruzzelli, A. D'Orazio, F. Prudenzano, D. de Ceglia, N. Akozbek, M.J. Bloemer, P. Ashley, and M. Scalora, "Enhanced transmission and second harmonic generation from subwavelength slits on metal substrates", *Proc. SPIE* **6987**, 69870O (2008).
42. J. A. Porto, F. J. Garcia-Vidal, and J. B. Pendry, "Transmission resonances on metallic gratings with very narrow slits," Phys. Rev. Lett. **83**, 2845 (1999).
43. Q. Cao and Ph. Lalanne, "Negative Role of Surface Plasmons in the Transmission of Metallic Gratings with Very Narrow Slits," Phys. Rev. Lett. **88**, 057403 (2002).
44. P. Lalanne, C. Sauvan, J. P. Hugonin, J. C. Rodier, and P. Chavel, "Perturbative approach for surface plasmon effects on flat interfaces periodically corrugated by subwavelength apertures," Phys. Rev. B **68**, 125404 (2003).
45. Y. Xie, A. R. Zakharian, J. V. Moloney, and M. Mansuripur, "Transmission of light through a periodic array of slits in a thick metallic film," Opt. Express **13**, 4485 (2005).
46. D. Pacifici, H. J. Lezec, H. A. Atwater, and J. Weiner, "Quantitative Determination of Optical Transmission through Subwavelength Slit Arrays in Ag films: The Essential role of Surface Wave Interference and Local Coupling between Adjacent Slits," Phys. Rev. B **77**, 115411(2008).
47. R. W. Boyd, *Nonlinear Optics,* Academic Press, New York (2003)
48. V. Roppo, F. Raineri, C. Cojocaru, R. Raj, J. Trull, I. Sagnes, R. Vilaseca, M. Scalora, "Generation efficiency of the Second Harmonic Inhomogeneous Component," arXiv:1010.4693v1, (2010).
49. R. S. Bennink, Y. Yoon, R.W. Boyd, and J. E. Sipe, "Accessing the optical nonlinearity of metals with metal- dielectric photonic bandgap structures,"Opt. Lett. **24**, 1416 (1999).
50. J. Olivares, J. Requejo-Isidro, R. del Coso, R. de Nalda, J. Solis, C. N. Afonso, A. L. Stepanov, D. Hole, P. D. Townsend, and A. Naudon, "Large enhancement of the third-order optical susceptibility in Cu-silica composites produced by low-energy high-current ion implantation ,"J. Appl. Phys. **90**, 1064 (2001).
51. J. M. Ballesteros, R. Serna, J. Soli's, C. N. Afonso, A. K. Petford-Long, D. H. Osborne, and R. F. Haglund, "Pulsed laser deposition of Cu:Al2O3 nanocrystal thin films with high third-order optical susceptibility,"Appl. Phys. Lett. **71**, 2445 (1997).